\documentstyle[11pt,aaspp4,flushrt,psfig]{article}

\begin{document}
\pagestyle{plain}

\title{The Extinction Towards the GRB970228 Field}

\author{Francisco J. Castander and Donald Q. Lamb}
\affil{Department of Astronomy and Astrophysics, University of Chicago,\\
    5640 S Ellis Ave, Chicago, IL 60637}





\begin{abstract}

We determine the local galactic extinction towards the field of gamma-ray
burst GRB970228 using a variety of methods.  We develop a maximum
likelihood method for measuring the extinction by comparing galaxy counts
in the field of interest to those in a field of known extinction, and apply
this method to the GRB970228 field.  We also measure the extinction by
comparing the observed stellar spectral energy distributions of stars in
the GRB970228 field to the spectral energy distribution of library spectra
of the same spectral type. Finally we estimate the extinction using the
Balmer emission line ratios of a galaxy in the GRB970228 field, and the
neutral hydrogen column density and amount of infrared dust emission toward
this field.  Combining the results of these methods, we find a best-fit
galactic extinction in the optical of $A_V=1.19^{+0.10}_{-0.17}$, which
implies a a substantial dimming and change of the spectral slope of the
intrinsic GRB970228 afterglow.

\end{abstract}

\keywords{gamma rays: bursts --- extinction}


%

\section{Introduction}

GRB970228 is the first gamma-ray burst (GRB) for which a counterpart at
longer wavelengths has been detected, and extensive follow-up observations
of it have been made. GRB970228 was detected by the BeppoSAX satellite on
1997 February 28 (\cite{cos97a}). Subsequent BeppoSAX follow-up
observations revealed a rapidly fading X-ray source (\cite{cos97b}).
Later, ASCA (\cite{yos97}) and ROSAT (\cite{fro97}, 1998) observations
showed that it continued to fade in X-rays over a two week period.

Ten days after the burst, Groot et al. (1997a)\markcite{gro97a} announced
the detection of a fading source which was the first optical counterpart of
a GRB. Frenetic activity followed, with new observations being taken and
previous ones being reanalyzed, which led to reports of several detections
in the optical and near infrared (\cite{gro97b}; \cite{met97a};
\cite{vpa97}; \cite{met97b}; \cite{klo97}; \cite{mar97}; \cite{soi97};
\cite{met97b}; \cite{ped97}; \cite{djo97}).  Observations with the Hubble
Space Telescope made another startling discovery. The fading GRB afterglow
was spatially coincident with an extended source (\cite{sah97a};
\cite{sah97b}; \cite{fru97}).

\markcite{wij97}Wijers et al. (1997) and \markcite{rei97}Reichart
(1997) discussed early observations of GRB970228 in the context of
theoretical models. They found that the behavior of the GRB afterglow
was consistent with the expectations of relativistic fireball models.
In one of the simplest models, the afterglow power-law temporal decay
is simply related to its power-law spectrum by a factor of 1.5, which
was consistent with the early measurements that were being reported.
Later, \markcite{gal97}Galama et al. (1997, 1998) compiled the most
relevant photometric measurements of GRB970228, converting them into a
single photometric band when necessary and subtracting the contribution
of the extended source component. They fit the optical transient
temporal evolution to a power-law with $\alpha=-1.10\pm0.04$
($\chi^2_r=2.3$ for 9 degrees of freedom).

For the first time, temporal, as well as spatial, coincidence could be used
to associate X-ray and optical sources with GRBs. Since then several other
GRB afterglows have been detected and monitored, and it now seems firmly
established that these fading X-ray and optical sources are counterparts of
the bursts. In this paper, we determine the galactic extinction towards
GRB970228, performing a careful analysis of the publicly available
observations. The goal we have in mind is to better understand the
intrinsic properties of this afterglow. In a companion paper (\cite{CL98}),
we discuss the implications of our measurements for the properties and
nature of the point-like and extended optical sources coincident with the
fading X-ray source.

This paper is organized as follows. First, we utilize various methods
to measure the extinction at optical wavelengths. In \S2 we describe a
photometric method: we compute the galaxy number counts in the
GRB970228 HST WFPC2 observations and compare them to the number counts
in the Hubble Deep Field (HDF; \cite{wil96}). In the next section we
describe spectroscopic methods that use the Keck II observations made
by \markcite{ton97}Tonry et al. (1997). In \S3.1 we measure the Balmer
series emission lines of a galaxy 64'' away from the GRB and
estimate the extinction from the observed relative intensities of these
emission lines. In \S3.2 we use three stars that lie 2.9'', 16.8'' and
42.7'' away from the GRB.  We determine their spectral types and
compute the extinction value required to make their measured spectral
energy distributions consistent with their spectral types. Secondly, we
estimate the extinction from measurements of the column density of
hydrogen gas and dust emission measurements, using established
correlations. In \S4 we utilize hydrogen column density measures and in
\S5 the infrared 100 {\micron} dust emission. We discuss our results in
\S6 and present our conclusions in \S7.

\section{Galaxy number counts}

Galaxy number counts can be used to measure directly the relative
extinction between two fields. The idea is simple. The observed apparent
optical magnitude of a galaxy is increased (the flux is decreased) because
its radiation is absorbed by material, normally dust for optical
extinction, along the line of sight. Because the number of observed
galaxies increases with magnitude, number counts are reduced if extinction
is present. Ignoring possible deviations due to galaxy clustering and
sampling effects, in a given magnitude range and in a given filter, the
number of galaxies should be the same irrespective of direction. Galaxy
number counts are normally approximated as
 
\begin{equation}
N(m_1<m<m_2) = C 10^{\alpha m},
\end{equation}
where the values of the normalization, $C$, and the slope, $\alpha$, depend
on the specific filter and magnitude range. However, due to extinction, the
observed apparent magnitude will be increased to $m_{obs}=m + A$.
Therefore, if we compare two different fields with extinctions $A_1$ and
$A_2$, their relative number counts in the same observed apparent magnitude
range will be 

\begin{equation}
\frac{N_1(m_1<m_{obs}<m_2)}{N_2(m_1<m_{obs}<m_2)} = 
\frac{C_1}{C_2} 10^{\alpha_1 (m+A_1) - \alpha_2 (m+A_2)}.
\end{equation}

One can assume that the normalizations and slopes are the same, if the
galaxy number counts are measured in the same unextinguished apparent
magnitude range and in the same filter. If we additionally assume that
surface brightness dimming effects do not alter the relative number counts,
then the ratio of the number counts depends only on the relative extinction
and common slope

\begin{equation}
\frac{N_1(m_1+A_1<m_{obs}<m_2+A_1)}{N_2(m_1+A_2<m_{obs}<m_2+A_2)} = 
10^{\alpha (A_2-A_1)}.
\end{equation}

In the present case we wish to estimate the extinction towards GRB970228 by
comparing the number counts of the GRB WFPC2 HST observations with those of
another field of known extinction. We have chosen the Hubble Deep Field
(\cite{wil96}) because it is the best studied and deepest field for which
the extinction is already known, having been observed with the same
instrument.

GRB970228 was observed by HST on 1997 March 26th and April 7th. In both
observations the optical counterpart was centered in the middle of the PC1
CCD, but there was a 2.40 deg difference in rotation angle between the
first and second observations. At both epochs, four exposures were taken in
the F606W filter and two exposures in the F814W filter, totalling 4700 and
2400 seconds, respectively (\cite{sah97c}). The HDF was observed for 109050
and 123600 seconds in the F606W and F814W filters respectively (for more
details see \cite{wil96}).

Our starting point was the HST archive, from which we retrieved the
observations of both fields. After the standard pipeline reduction, we
combined the different exposures at each epoch of the GRB970228 field using
the IRAF/STSDAS task CRREJ. Then the combined second epoch image was
rotated according to the difference in the ORIENTAT header keyword and
shifted using sub-pixel shifts and fourth-order polynomial interpolation,
according to the measured centroid positions of stars. Subsequently, both
epochs were combined using the IRAF task IMCOMBINE, rejecting pixels that
were deviant from the median by more than 5 sigma, the noise being
characterized by the square root of the median plus the square of the
read-out noise. However, for this last step to be effective the sky values
had to be rescaled because the direction of the April 7th observations was
closer to the Sun than was the direction of the previous ones, and the
images had background count values approximately 25\% higher. This last
rejection process affected only $\sim$0.12\% and $\sim$0.35\% of the pixels
in the F606W and F814W images, respectively. The HDF images retrieved had
already been processed and no further reduction was done.

In order to reduce the systematic errors in selecting galaxies and
measuring their magnitudes, we chose to analyze both fields using the same
procedures.  We used the SExtractor image analysis package (\cite{BA96}) to
automatically detect and measure object magnitudes.  The magnitude zero
points were computed using the PHOTFLAM and PHOTZPT image header keywords,
taking into account the different effective gains of the four WFPC2 CCDs
(\cite{hol95}).  These ST magnitudes were converted to AB
magnitudes\footnote[1]{All magnitudes quoted in this paper are in the AB
magnitude system} using the transformations: $(V_{606})_{AB} =
(V_{606})_{ST} - 0.199$ and $(I_{814})_{AB} = (I_{814})_{ST} - 0.819$.
After generating catalogs of extracted objects and their magnitudes, all
objects brighter than $V_{606}=26.5$ and $I_{814}=25.9$ were visually
inspected and spurious objects were removed. This process was crucial in
the F814W image of the GRB970228 field because combining only four
exposures precluded an accurate rejection of ``hot pixels''. The different
rotation angles between the March 26 and April 7 images of the GRB970228
field and the small dithering in the HDF made it necessary to exclude the
edges of the CCD fields as well. After generating the catalogs, we compared
our galaxy number counts in the HDF field with those obtained by the HDF
team, in order to check our object selection criteria and our magnitude
measurements. The galaxy number counts in both filter images were
consistent with each other.

In order to estimate the relative extinction between the two fields, we
utilize a maximum likelihood method. We construct a joint likelihood
function that is the product of four likelihoods, each one being the
likelihood that a given galaxy catalogue with its measured magnitudes and
errors resembles a power-law distribution (see equation~1) in a given
magnitude range. The four likelihood functions correspond to $V_{606}$ and
$I_{814}$ images of the two fields. Therefore our joint likelihood function
has 8 parameters: the normalizations and slopes in each of the four
combinations of images and filters ($C_{606}^{HDF}$, $\alpha_{606}^{HDF}$,
$C_{606}^{GRB}$, $\alpha_{606}^{GRB}$, $C_{814}^{HDF}$,
$\alpha_{814}^{HDF}$, $C_{814}^{GRB}$, $\alpha_{814}^{GRB}$). However, if
we analyze the catalogues in the same unextinguished apparent magnitude
range (which requires us to know the extinction a priori; see equation~3),
we can assume that the slopes are the same in the HDF and GRB970228 fields
in a given filter image, and that the normalizations are therefore related
by

\begin{displaymath}
\alpha_{606}^{GRB} = \alpha_{606}^{HDF}, 
\end{displaymath}
\begin{displaymath}
\alpha_{814}^{GRB} = \alpha_{814}^{HDF},
\end{displaymath}
\begin{displaymath}
C_{606}^{GRB} = C_{606}^{HDF} \:10^{\alpha  
(A_{606}^{HDF}-A_{606}^{GRB}) },
\end{displaymath}
\begin{equation}
C_{814}^{GRB} = C_{814}^{HDF}\: 10^{\alpha  
(A_{814}^{HDF}-A_{814}^{GRB})}.
\end{equation}

If we further assume that the extinction behaves like extinction law
typical of the interstellar medium (\cite{car87}, \cite{odo94}), we can
impose a further constraint by integrating the extinction law with the
filter responses

\begin{displaymath}
A_{606} = 0.919 \: A_V,
\end{displaymath}
\begin{equation}
A_{814} = 0.608 \: A_V.
\end{equation}

Then the joint likelihood function has only 5 parameters: two slopes, two
normalizations, and the difference in the extinction. We can marginalize
this likelihood function over both normalizations (F606W and F814W) and use
the measured value of the extinction in the HDF field, $A_V=0.0$
(\cite{wil96}), in order to reduce the maximization required to three
parameters ($\alpha_{606}$, $\alpha_{814}$, $A_V$).

\begin{figure}[t]
\columnwidth9cm
\plotone{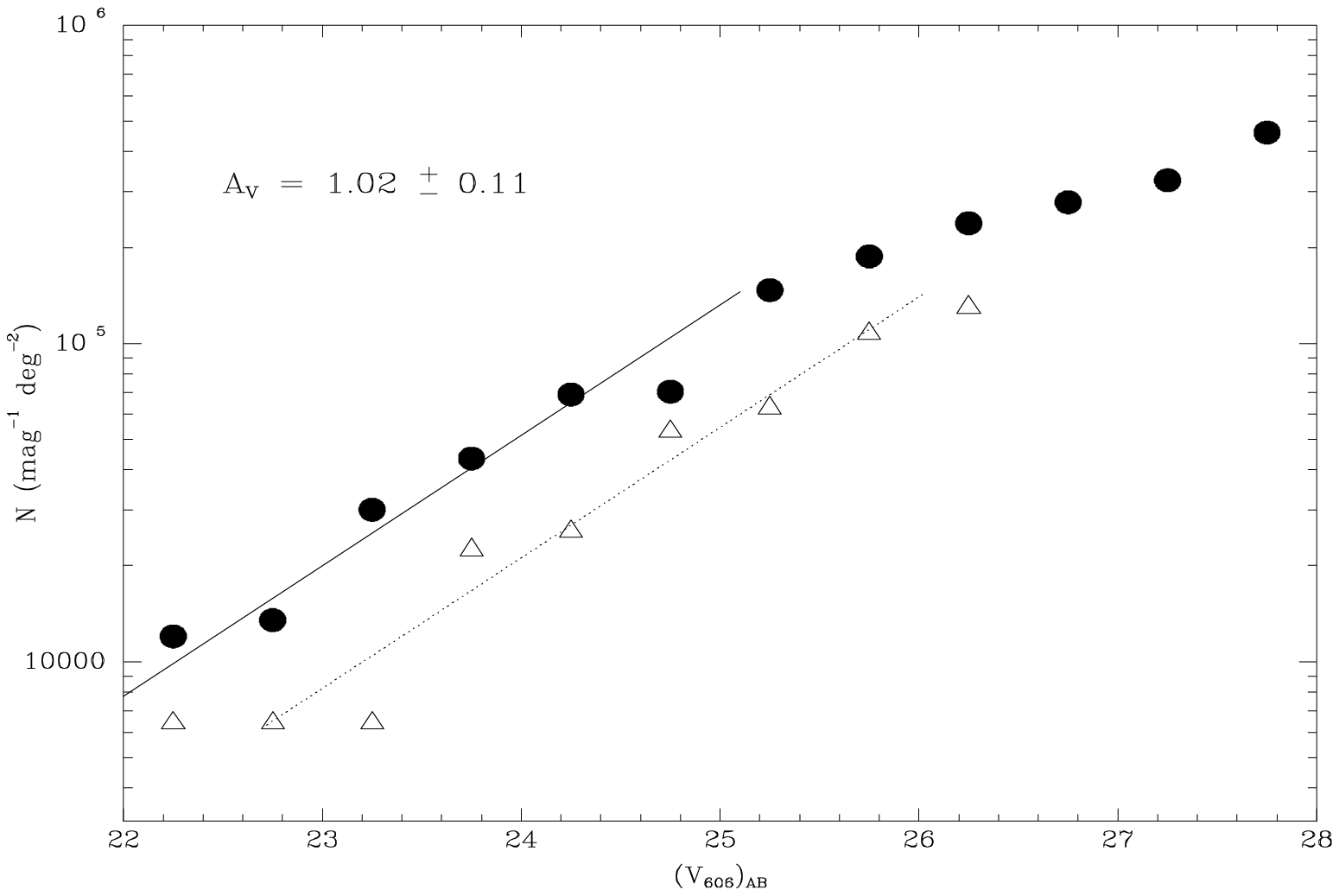}\\
\plotone{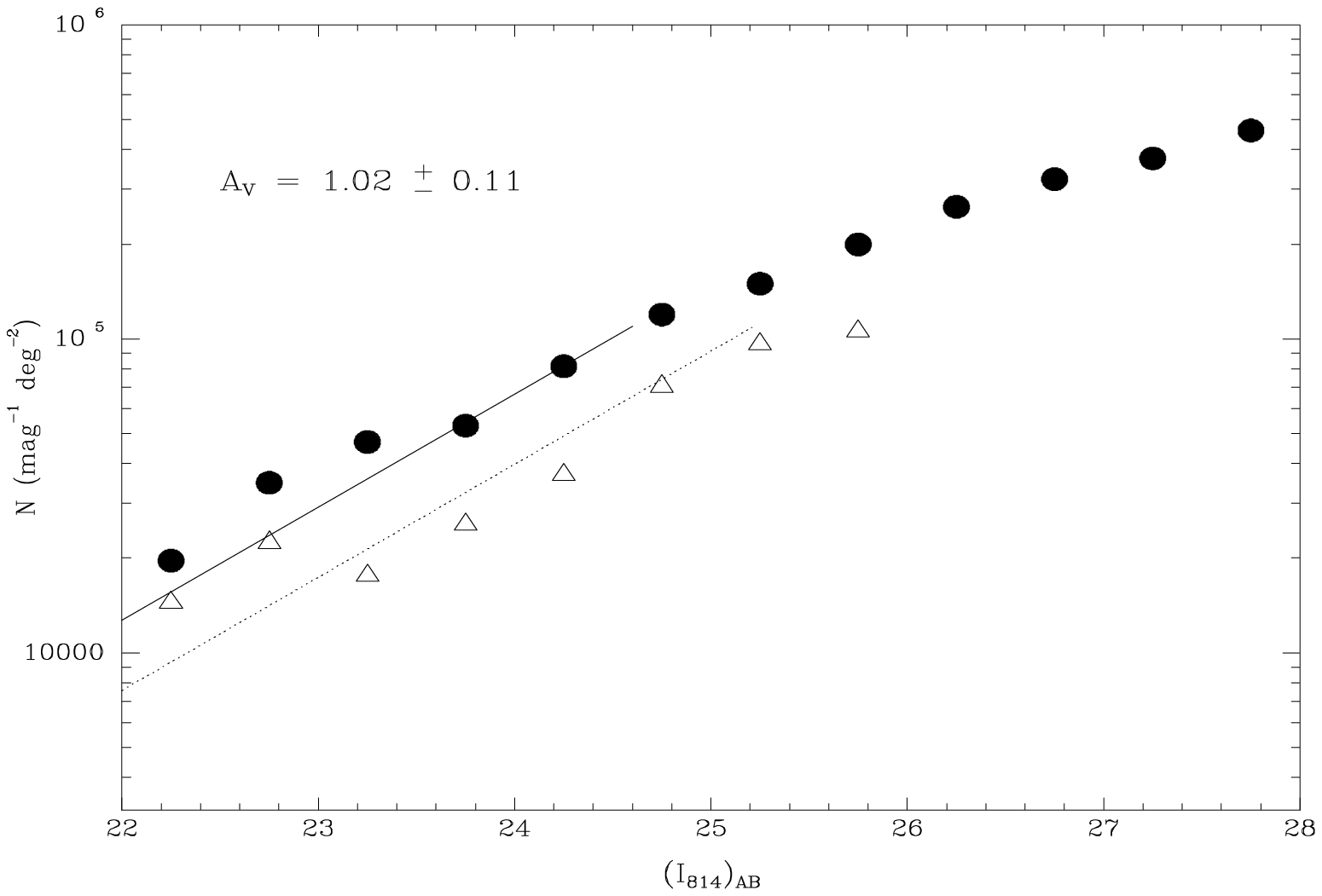}
\caption[]
{Number counts in the HDF (filled circles) and GRB970228 (open triangles)
fields. The top and bottom figures correspond to the $V_{606}$ and
$I_{814}$ filters respectively. Error bars are omitted for clarity. The
solid and dotted lines represent the best fit slope values in the
combined maximum likelihood method. They are plotted only in the regions
were the fit was performed. Note that they are not independent (see
text for details).
}
\label{fig1}
\end{figure}

As mentioned before, we need to know {\it a priori} the value of the
extinction in order to choose the magnitude ranges for which our parameter
reduction is valid. Therefore, we proceed in the following iterative
way. We start by choosing a value for the extinction (a good initial guess
can be made by comparing the counts in different magnitudes ranges). We
then correct the observed magnitudes for that extinction value and plot the
resulting cumulative number counts distributions.  Since the HDF field has
longer exposures and goes much deeper, we can compute the
extinction-corrected magnitude at which the GRB970228 field counts start to
be incomplete, compared to the HDF field. Plotting the cumulative number
counts of both extinction-corrected datasets gives a good indication where
the incompleteness starts to affect the GRB970228 field. We adopt as our
faint limiting magnitude the magnitude at which the cumulative counts
deviate by more than 10\% in the F606W filter and 15\% in the F814W filter
(the errors in the computed magnitudes are larger in the F814W filter and
that is why the deviation allowed is larger as well). Note that the
deviation in the F814W filter depends on the bright magnitude limit adopted
as the HSD and GRB counts differ at the lower $I_{814}$ magnitudes measured
(see Figure~\ref{fig1}). We also check the probability as a function of
limiting magnitude that both cumulative magnitude distributions are drawn
from the same parent distributions, using the KS test. At our adopted cut and
taking into account our bright magnitude limit as well, the hypothesis that
the two count distributions come from the same parent distribution cannot
be rejected at even the $1\sigma$ confidence level. Once we have determined
the limiting extinction-corrected magnitude, we use it, without the
extinction correction, as the faintest magnitude down to which we compute
the joint likelihood function. We take as the bright magnitude limit the
magnitude at which the number counts fall below $\sim 7.5 \times 10^4$
mag$^{-1}$ deg$^{-2}$ ($\sim$10 objects per magnitude in the whole WFPC2
area).  We then maximize the likelihood function and obtain the best-fit
values for the slopes and the extinction. Starting with this new value for
the extinction, we iterate until the process converges.

It is worth noting that our final value for the extinction is almost
independent of the bright magnitude limit adopted.  The exact faint
magnitude limit adopted does not affect the final value of extinction
either, because we perform a correction for possible incompleteness {\it a
posteriori} (see below).

Figure~1 illustrates our results. After a few iterations, we obtain values
for the extinction and number counts slopes of $A_V=1.02\pm0.11$,
$\alpha_{606}=0.41\pm0.1$, and $\alpha_{814}=0.36\pm0.1$. The magnitude
ranges used are $21.8 < V_{606}^{HDF} < 25.1$, $22.72 < V_{606}^{GRB} <
26.02$, $21.0 < I_{814}^{HDF} < 24.6$, and $21.61 < I_{814}^{GRB} < 25.21$.

So far we have neglected any possible effects introduced by the dimming of
the surface brightnesses of galaxies due to the optical extinction. In the
present case, the GRB field is considerably obscured compared to the HDF
and therefore low surface brightness objects could be missed in the GRB
field because they fall below the detection limit when they should have
been included in the magnitude limits considered. This effect would lead to
an overestimation of the extinction. In order to correct for it, we
simulate WFPC2 fields by degrading the WFPC2 HDF images according to
several values of the extinction and then carry out the same analysis on
these simulated images as we have on the GRB field images. According to our
simulations, we overestimate the value of the extinction by approximately
15\%.

Our extinction estimate could also be affected by clustering if one of the
fields we have chosen to study is more or less strongly clustered compared
to the other.  The integral of the two point angular correlation function
gives an estimate of the extra variance introduced by clustering in the
number counts. In the magnitude range studied here, the two point angular
correlation function can be approximated by $\omega(\theta\arcsec)\sim 1.0
\,\theta\arcsec^{-0.8}$ (\cite{BI98}). Integrating over the WFC2 area
studied, we find that the angular clustering on the sky adds a small
contribution to the variance in the counts at the magnitude range studied.
If we include this additional variance contribution into our maximum
likelihood method and also take into account the previous $\sim$15\%
correction factor, we obtain a revised extinction value of
$A_V=0.89\pm0.13$, where the estimated error includes the contribution of
our correction factor as well as the contribution resulting from our {\it
clustering-modified} maximum likelihood method.

\section{Spectroscopic methods}

Spectroscopy of the GRB optical afterglow was attempted using the Keck
II telescope by \markcite{ton97}Tonry et al. (1997).  Although they
could not obtain a spectrum with sufficient signal-to-noise to discern
the nature of the optical counterpart, they obtained spectra of several
other nearby objects that fell within the long slit they used (see
Figure~\ref{fig2}).

\begin{figure}[t]
\centerline{
Figure available at:
}
\centerline{
http://astro.uchicago.edu/home/web/fjc/figures/CL98a/CL98a.fig2.eps
}
\caption[] 
{Slit position of the Keck II spectroscopic observations of the GRB970228
field (\cite{ton97}). The distance from S2 to S3 is 59.5''.
}
\label{fig2}
\end{figure}

We retrieved the spectroscopic data from the Hawaii public FTP directory
and reduced them. Observations had been taken with the Low Resolution
Imaging Spectrograph (LRIS) in six exposures on March 31.25 UT (500 s and
1000 s), April 1.25 UT (1000 s and 1000 s) and April 2.25 UT (1000 s and
1000 s).  The instrument configuration used was the 300 lines/mm grating,
blazing at 5000 \AA, giving a dispersion of $\sim$2.5 \AA/pixel and an
approximate wavelength coverage 4300-9500 {\AA}. The slit was 1.0'' wide
and 2.9' long. It was centered at the star 2.9'' East of the optical
counterpart (S1) and moved 10'' Eastward in two of the exposures. The
position angle was 86.4 degrees.

The spectroscopic images were processed using standard IRAF routines. We
combined the observations into one image, shifting the offset images and
rejecting cosmic rays. The resulting optimally-extracted spectra were found
to be insensitive to the type of shift applied: fractional or integer pixel
shifts. Six objects fell, totally or partially, within the slit and
produced dispersed spectra in the 2-dimensional images. Figure~\ref{fig2}
shows where the slit was placed in the sky and the objects for which
spectra were obtained. We did not include in the final combined image the
observations taken on March 31.25 UT, due to their poorer quality. The total
exposure used was then 4000 s.

We extracted the spectra with our own implementation of Horne's optimal
extraction algorithm (\cite{hor86}). Spectra were wavelength-calibrated
with He-Ne arcs and flux calibrated with the spectrophotometric standard
Hiltner 600. We obtained spectrophotometric colors convolving the flux
calibrated spectra with the WFPC2 filter responses. Comparing these to the
photometric colors measured in the WFPC2 HST image we found our relative
flux calibration errors to be lower than 5\% (see Table~\ref{tbl1}).

\subsection{Balmer series emission line ratios of galaxies}

The flux ratios of the Balmer emission lines of galaxies can be used to
characterize the optical extinction.  The $H_{\alpha} / H_{\beta}$ and
$H_{\gamma} / H_{\beta}$ ratios of emission-line galaxies depend on the
conditions within that galaxy. Theoretically, for typical galaxy
conditions, these values are 2.88 and 0.46 respectively (\cite{ost89}).
The measured values, however, will differ due to the intrinsic extinction
within the galaxy at redshift $z$ and to the local extinction in our own
galaxy. Once the Balmer line flux ratios are measured, and assuming a
typical extinction law, one can compute both the intrinsic and the local
extinction values.

\begin{figure}[t]
\columnwidth12cm
\plotone{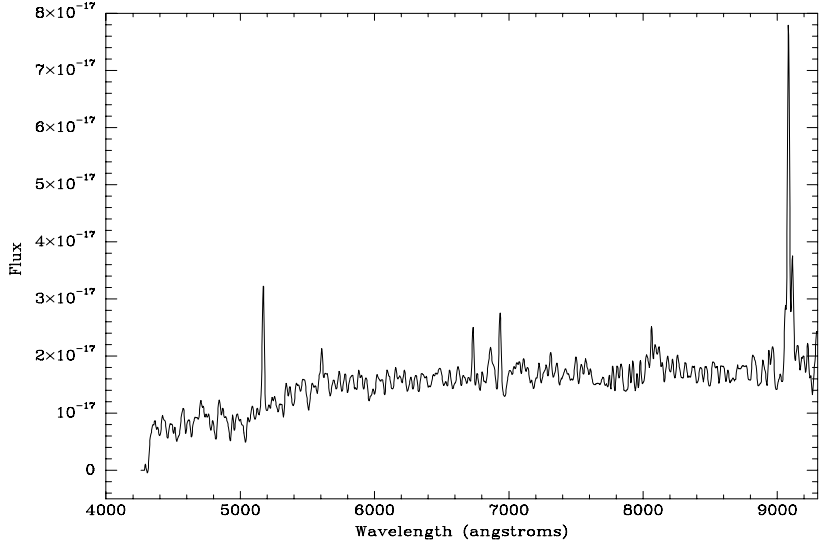}
\caption[]
{Slightly smoothed spectrum of galaxy G2 at redshift $z=0.3792$.
}
\label{fig3}
\end{figure}

\begin{figure}[t]
\columnwidth12cm
\plotone{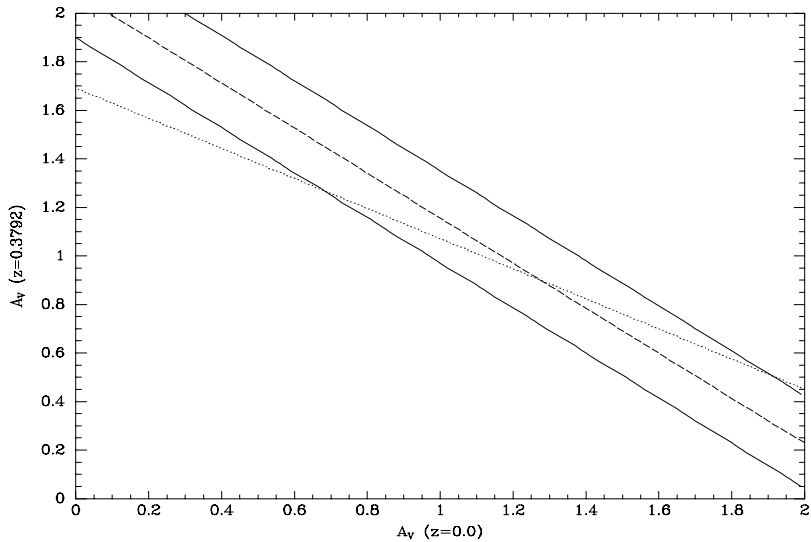}
\caption[]
{Balmer ratio constraints on the internal extinction within galaxy G2 at
$z=0.3792$ and the extinction due to our own galaxy. The constraint for a
ratio is degenerate in this extinction-extinction plane. The dashed line
gives the best fit extinction values based on the $H_{\alpha}/H_{\beta}$
ratio and the solid line its $1\sigma$ error. The dotted line, the
$H_{\beta}/H_{\gamma}$ ratio constraint. Due to the uncertainty of this
ratio measurement no $1\sigma$ errors are plotted. The best values obtained
are $A_V (z=0.3792) = 0.901$ and $A_V (z=0.0) = 1.276$. However these
values are only constrained by the $H_{\alpha}/H_{\beta}$ ratio region.
}
\label{fig4}
\end{figure}

In the Keck II spectroscopic observations, the two galaxies that happened
to lie within the slit exhibited emission lines. However, only the spectrum
of the galaxy~G2, at redshift $z=0.3792$, had a large enough
signal-to-noise ratio to allow reliable measurements of the Balmer emission
line fluxes (Figure~\ref{fig3}). We measure the following Balmer line flux
ratios: $H_{\alpha}/H_{\beta} = 5.60^{+0.36}_{-0.32}$ and
$H_{\beta}/H_{\gamma} = 0.35^{+0.12}_{-0.18}$, where the errors are mainly
due to the uncertaintanty in estimating the continuum. Assuming a standard
value for these ratios of 2.88 and 0.465 (\cite{ost89}) and a typical
extinction law for diffuse interstellar medium, $R_V=3.1$ (\cite{car87}),
we can obtain the values for the extinction intrinsic to the galaxy and
within our own galaxy. Figure~\ref{fig4} shows the values allowed with our
measured ratios and assumptions. Our best values are: $A_V (z=0.3792) =
0.901$ and $A_V (z=0.0) = 1.276$. However, due to the uncertainties in the
ratios and the almost parallel constraints that the ratios place in the $A_V
(z=0.3792)$ vs. $A_V (z=0.0)$ plane, a wide range of values are allowed
(Figure~\ref{fig4}).

\subsection{Stellar spectral energy distributions versus spectral types}

Another method that can be used to estimate the extinction in this field is
to compare the spectral energy distributions and spectral types of stars
observed spectroscopically. If the spectral type can be reliably determined
without color or spectral continuum information, then one can infer the
extinction by finding the best fit of the extinction-corrected spectrum
(either in color or continuum spectral distribution) to library spectra of
the same spectral type, with the extinction correction as a fitting
parameter.

As mentioned before, three stars were observed spectroscopically (two
accidentally) with the Keck II telescope by \markcite{ton97}Tonry et
al. (1997) (Figure~\ref{fig2}). The signal-to-noise ratios achieved in
the 6000-7000 {\AA} spectral region were 3.7, 11.5 and 1.3 for stars
S1, S2 and S3, respectively.

These stars were also observed within the WFPC2 HST pointings of the
GRB970228 field. Table~\ref{tbl1} summarizes the magnitudes and colors
measured for these stars.  The apparent magnitudes indicate that these
stars are most likely to be dwarfs. If they were giants, their apparent
magnitudes would imply that their distances are larger than 100 kpc in
order to be consistent with absolute magnitudes of giant stars of the same
spectral type. Such distances are highly unlikely.

In order to classify these stars, we compare them with the stellar spectral
atlases of \markcite{jac84}Jacoby et al. (1984) and \markcite{SC92}Silva \&
cornell (1992). Given the wavelength coverage and the low signal-to-noise
ratio of the stellar spectra, there are only a few features that can be
used to classify these stars. The wavelength coverage does not allow us to
use features bluewards of $H_{\beta}$. We classify S1 in a spectral class
between K3v and K5v, S2 between K4v and K7v and S3 between M0v and M3v. The
K-star classifications are mainly based on the Mg $\lambda$5174 and Na
$\lambda$5893 $W_f$ indexes (\cite{PB77}), and on the presence or absence
of TiO, VO and CaH bands (e.g., \cite{kir91}).

We have also attempted a more automatic classification, based only on line
information. We fit and subtracted a continuum to the observed stars and to
the template spectra. We then computed a metric distance (treating the
spectra as vectors; e.g., \cite{VP95}) between the observed spectra and the
templates. We also computed a standard $\chi^2$ between the observed
continuum-subtracted spectra and the continuum-subtracted library
spectra. Both methods produce similar results. The spectral types that give
the minimum metric distances and values of $\chi^2$ for our stars are: S1,
from K1v to K7v; S2, from K2v to K5v and S3, from M0v to M4v, although the
constraints for S3 are rather weak.

\begin{figure}[t]
\columnwidth9cm
\plotone{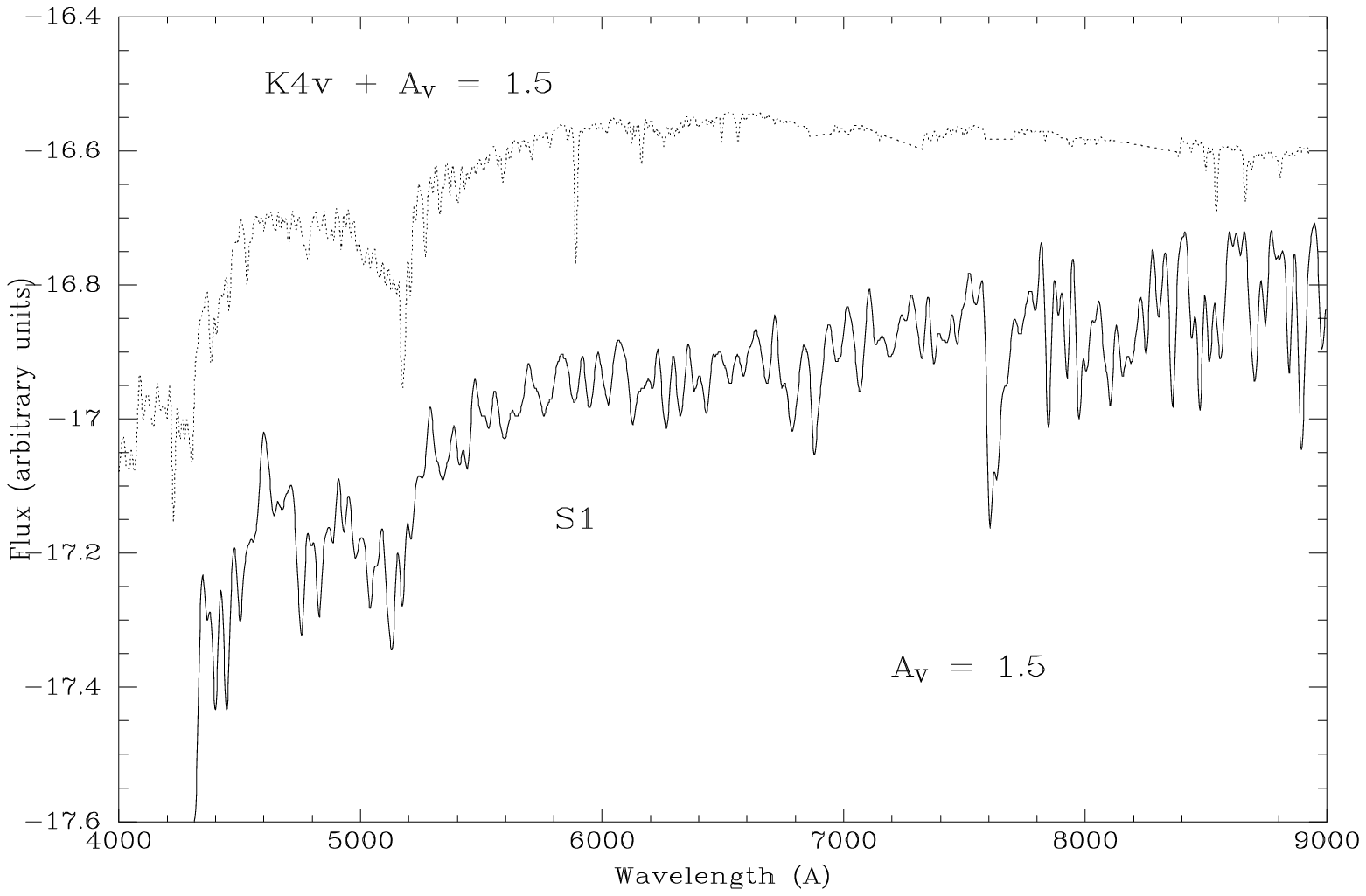}\\
\plotone{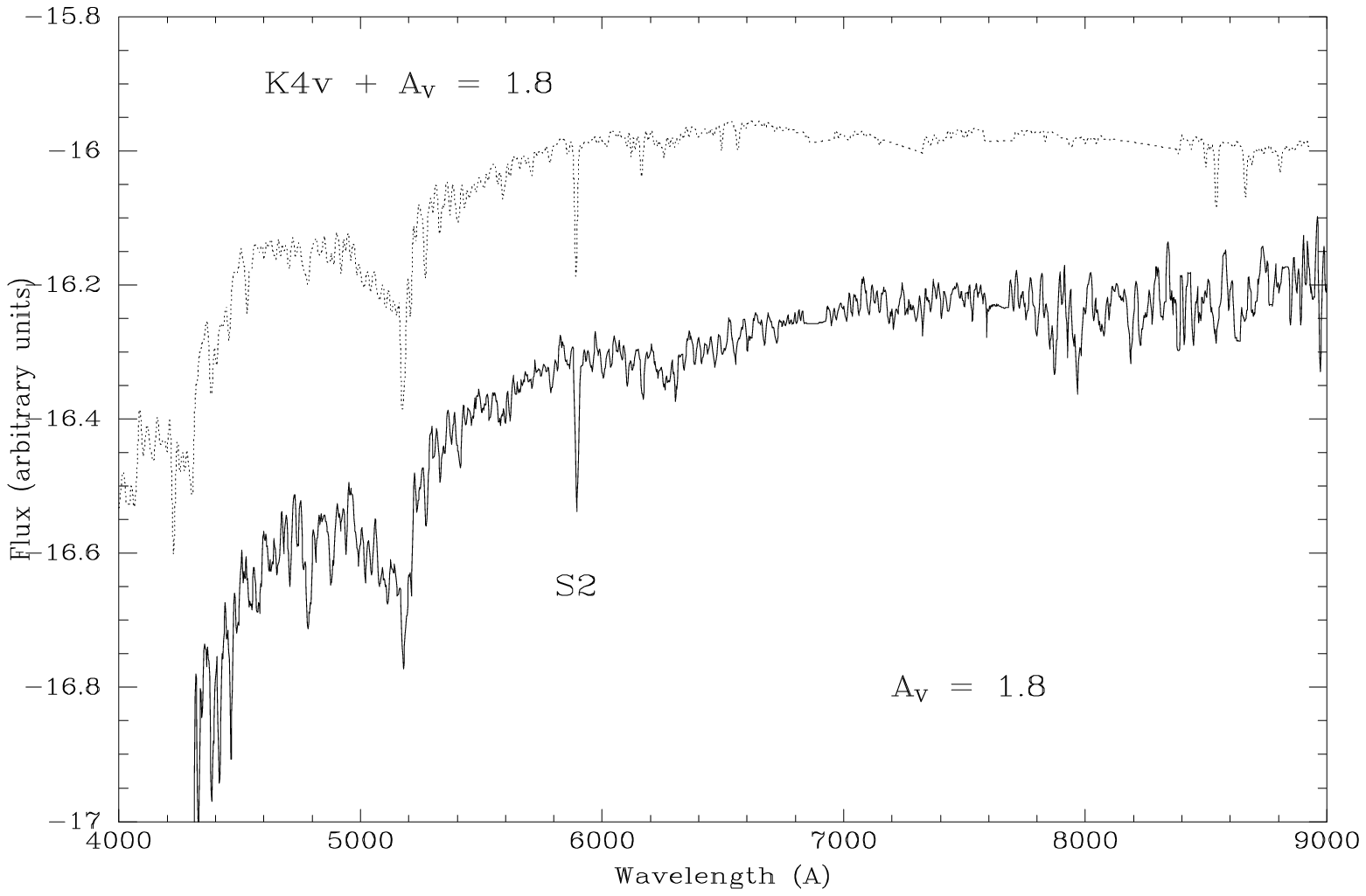}
\caption[]
{Spectra of stars S1 (top) and S2 (bottom). The solid lines correspond to
the observed star spectra, while the dotted lines are the library spectra
reddened to match the observed spectra (see text for details).
}
\label{fig5}
\end{figure}

Taking into consideration our visual classification, the measured spectral
indexes and the automatic classification allowed types, we estimate the
value of the extinction for each star by reddening the template spectra to
fit the spectral distribution of our observed stars. In order to avoid
possible problems with sky subtraction, we have used only the wavelength
region from $4500-7500$ {\AA} in the fitting procedure. Figure~\ref{fig5}
illustrates the comparison between the observed and reddened library
spectra.

For S1, we obtained $A_V=1.5^{+0.4}_{-0.9}$ with the best value
corresponding to a K4v star and the allowed range corresponding to K1v and
K7v stars.  For S2, we get $A_V=1.8^{+0.2}_{-0.5}$ with the best value
corresponding to a K4v star, and the allowed range to a K2v star and an
intermediate K5v-K7v star.  For S3, $A_V=1.8^{+0.5}_{-1.0}$, the best value
corresponds to a M2v star and the allowed range to M0v and M3v stars. The
errors quoted include in quadrature the uncertainty in the stellar spectral
fit, given by the allowed ranges of spectral types, and an estimate of the
uncertainty contributed by the spectrophotometric calibration.

Finally, we compute our best estimate of the extinction by combining the
contributions of the three stars weighting them by their signal-to-noise.
We obtain a value of $A_V=1.73^{+0.20}_{-0.42}$.

\section{Extinction from neutral hydrogen column density}

BeppoSAX detected the GRB970228 afterglow in X-rays with both the MECS and
LECS instruments (\cite{cos97b}). In the first set of observations, eight
hours after the burst, the measured flux was bright enough to fit a spectrum
to the measured counts in the 0.1-10 keV energy band. The best fit was
obtained for a power-law spectrum with photoeletric absorption of
$N_H=3.5^{+3.3}_{-2.3} \times 10^{21}$ cm$^{-2}$ (\cite{fro98}). Converting
this value to color excess (see below), and assuming that all of the
absorption is due to our own galaxy, would imply a color excess of $E(B-V)=
0.73^{+0.69}_{-0.48}$ and an optical extinction of
$A_V=2.26^{+2.13}_{-1.48}$.

Searching published HI surveys, we find values for the neutral hydrogen
column density of $1.60 \times 10^{21}$ and $1.59 \times 10^{21}$ cm$^{-2}$
from \markcite{sta92}Stark et al. (1992) and \markcite{DL90}Dickey \&
Lockman (1990), respectively. Adopting the conversion factor between HI and
color excess of $4.8 \times 10^{21}$ cm$^{-2}$ mag$^{-1}$ (\cite{boh78}),
we obtain $E(B-V)= 0.33$. This conversion factor is almost the same as that
obtained by \markcite{hei76}Heiles (1976) and somewhat lower than that of
\markcite{KK74}Knapp \& Kerr (1974).  According to \markcite{boh78}Bohlin
et al. (1978), although they do not quote a formal error, their conversion
factor should be accurate to within a factor of 1.5 for the total
(HI+H$_2$) hydrogen column density.  Comparing their total and neutral
hydrogen plots against color excess, the neutral hydrogen column density
shows a larger dispersion.  Nevertheless, adopting a factor of 1.5 would
give a color excess of $E(B-V)= 0.33\pm0.13$. If instead we adopt the
calibration of Heiles (1976)\markcite{hei76} for which the author gives
errors, a value of neutral hydrogen column density of $1.60 \times 10^{21}$
cm$^{-2}$ would yield a color excess of $E(B-V)= 0.29\pm0.03$. The
difference between these two values is due to a zero point difference in
the relation.

However, other methods of determining the color excess give slightly
different values. For example, \markcite{BH82}Burstein \& Heiles
(1982), using a combined HI column density/galaxy number counts method,
get a value of $E(B-V)= 0.23$, using the \markcite{hei75}Heiles (1975)
and \markcite{HC79}Heiles \& cleary (1979) HI measurements and the
smoothed \markcite{SW67}Shane \& Wirtanen (1967) galaxy counts.  If
instead of using the hydrogen column density value used by
\markcite{BH82}Burstein \& Heiles (1982), we use the
\markcite{sta92}Stark (1992) and \markcite{DL90}Dickey \& Lockman
(1990) value, adopting the conversion factor employed by
\markcite{BH82}Burstein \& Heiles (1982), and the value they give for
the smoothed galaxy counts from \markcite{SW67}Shane \& Wirtanen (1967)
we obtain $E(B-V)= 0.25\pm0.09$.

The error in the neutral hydrogen column density measurement is negligible
compared to the error in its relation to reddening, and therefore should
not add significant uncertainty to the color excess value derived. However,
the resolution of the HI maps is poor, around 1 or 2 degrees.  The
\cite{DL90} 21 cm HI map of the GRB970228 field is shown in
Figure~\ref{fig6}. As can be seen, the GRB970228 is in a region showing a
relatively steep HI gradient on angular scales of a few degrees and it is
quite conceivable that there are significant deviations on small angular
scales from the value assumed.

\begin{figure}[t]
\centerline{
Figure available at:
}
\centerline{
http://astro.uchicago.edu/home/web/fjc/figures/CL98a/CL98a.fig6.eps
}
\caption[]
{Hydrogen column density (left) and IRAS 100 micron (right) maps. Both are
$8.5\arcdeg \times 8.5\arcdeg$. Their resolutions are $\sim 1\arcdeg$ and
$\sim 5 \arcmin$, respectively. The white circle ($40 \arcmin$ diameter) is
centered on the position of the GRB970228 optical transient. The bright
regions correspond to strong emission, the dark regions to weak
emission. North is up and East to the left on both images. A strong
emission gradient is noticeable in the GRB970228 region.
}
\label{fig6}
\end{figure}

Taking into account these considerations, we conservatively adopt a value
of $E(B-V)= 0.30\pm0.13$. Assuming an extinction law typical of the diffuse
interstellar medium, $R_V\equiv A_V / E(B-V) = 3.1$ (\cite{car87},
\cite{odo94}), this gives an extinction value of $A_V=0.93\pm0.39$.

\section{Extinction from infrared emission}

The optical extinction is also known to correlate with infrared
emission at long wavelengths ($\sim 100 \mu m$). However, this
correlation shows substantial scatter because the infrared emission is
dependent on the radiation field and the temperature of the dust grains
(e.g., \cite{BP88}). The IRAS 100 $\micron$ emission map of the area is
shown in Figure~\ref{fig6}. The 100 $\micron$ infrared emission towards
GRB970228 is $I_{100_{\micron}}=13.1$ MJy sr$^{-1}$. \cite{row91} give
a conversion factor between visual extinction and dust emission of $A_V
/ I_{100_{\mu m}} = 0.06 $ mag / MJy sr$^{-1}$, that was computed
modeling the interstellar grains and their response to the interstellar
radiation field. This relation should be valid to within approximately
$\pm 30\%$. Thus we obtain a value for the extinction of $A_V=0.79\pm
0.21$. However, the GRB970228 field is located near the galactic
anticenter and towards that direction the intensity of the radiation
field declines and therefore the conversion factor is likely to be
higher than the value adopted, and consequently, the extinction value
underestimated.

Recently, \markcite{sch98}Schlegel et al. (1998) (hereafter SFD98) have
published reddening estimates based on COBE DIRBE and IRAS infrared
dust emission measures. They combine DIRBE data quality calibration
with IRAS resolution to get infrared measures, which they calibrate
using the color of elliptical galaxies and standard extinction curves
to obtain reddening estimates. They find a value for the GRB970228
direction of $A_V=0.70\pm 0.16$, consistent with
\markcite{BH82}Burstein \& Heiles (1982) and the IRAS 100$\micron$
emission with the \markcite{row91}Rowan=Robinson et al. (1991)
conversion factor.

\section{Discussion}

We have used several methods in order to determine the galactic extinction
towards the field of GRB970228. First, we have measured galaxy number
counts in the HST WFPC2 images and compared them to those in the HDF, a
field of known optical extinction. In order to reduce the systematic
errors, we have reanalyzed the HDF using the same techniques that we
utilized for the GRB field. Our object catalog for the HDF is very similar
to that of the HDF team, reinforcing the view that our selection method and
magnitude determinations are appropriate. We have also checked all selected
objects in the magnitude ranges studied to make sure that the different
depths of the images was not influencing the number counts, as could be the
case, for example, with the deblending algorithm. For instance, in the WF2
CCD image of the GRB970228 field there is a face on spiral that was over
deblended. Diffraction spikes also can produce automatic selection of
spurious objects. We corrected these manually.  This manual rejection
hardly affected our extinction estimate.  Another possible source of error
comes from star/galaxy separation. Again, because GRB field and the HDF
have different exposures, for the same magnitude ranges, objects in the HDF
are detected at a higher signal-to-noise ratio and can therefore be better
classified. Moreover, the GRB field is at a much lower galactic latitude,
increasing the surface number density of stars. Although the number of
stars detected and rejected is low, and therefore statistics on them are
poor, the number of stars detected is compatible with the expectations of
the Bahcall \& Soneira 1984\markcite{BS84} galactic model for this
particular galactic latitude. As a worst case scenario, we also estimated
the extinction with our maximum likelihood method, assuming that all
objects are galaxies. In this extreme case, the value of the extinction,
$A_V$, that we obtain is approximately 0.1 magnitudes lower than our best
estimate.

Another possible concern is the effect that extinction itself and the
different exposure times can have on the method. The GRB field shows
considerably more extinction than the HDF and was exposed for a much
shorter time. Incompleteness near the limiting magnitude of the GRB field
or low surface brightness objects being asigned fainter magnitudes or being
missed altogether could make us overestimate the extinction. We believe
that we have taken into account these effects with the correction factor we
derived from our simulations (see Section~2).

In our maximum likelihood analysis we have combined the F606W and F814W
images, assuming an extinction law. If we analyze both of the images
separately, we obtain consistent results when converting the extinction in
the WFPC2 filter to $A_V$, although the errors are larger. The F814W image
gives a somewhat higher estimate for $A_V$, as can be seen in
Figure~\ref{fig1}. The combined extinction measured, denoted by the
horizontal separation between the solid and dashed lines, seems
insufficient to explain the observed number counts. The combined extinction
value given by the maximum likelihood method is closer to the extinction
given by the F606W image than to the one given by the F814W image. The
F606W observations, having approximately double the exposure time, go
deeper and therefore more objects are detected in the observations taken in
this filter. The F606W counts therefore have slightly more weight in the
combined maximum likelihood method.

\markcite{gon98}Gonz\'alez et al. (1998) are also conducting an
investigation of the galactic extinction towards this same field. One of
their estimates comes from a comparison of the cumulative number counts in
the GRB field and the HDF in the F606W filter. In essence, their method is
very similar to our combined maximum likelihood method. It is therefore
reassuring that we obtain consistent values, within the errors (Fruchter,
private communication).

We have also measured the optical extinction by analyzing the spectra of
three nearby stars. Star S1 is the closest object to GRB970228, only 2.9''
away, and the other two are only 16.8'' (S2) and 42.7'' (S3) away. These
stars should then give the best estimate of the extinction if it varies on
small angular scales. Using stellar spectra has another advantage. A wide
wavelength range is sampled and the extinction signature can, in principle,
be better determined. Therefore we believe this method should provide the
best estimate of the optical extinction towards GRB970228. However, the
stellar spectra available are of poor signal-to-noise and do not cover blue
wavelength regions that are important for determining spectral types
(bluewards of $H_{\beta}$) and where the effects of extinction are
stronger. The main contribution in our error budget is then the allowed
range of spectral types. In comparison, the error due to the
spectrophotometric calibration is considerably smaller. Table~\ref{tbl1}
shows the consistency between the photometric and spectrophotometric
derived colors. This method, potentially superior to the others, is thus
somewhat hampered by the errors involved.

The Balmer lines ratios should also provide a powerful constraint on the
optical extinction. However, the two galaxy spectra taken by chance are of
low signal-to-noise, as the slit happened to fall far off the galaxy
centers. For galaxy G2 we are able to measure the $H_{\alpha}/H_{\beta}$
and $H_{\gamma}/H_{\beta}$ ratios. The value obtained from the former ratio
is degenerate in the internal galaxy extinction and the extinction due to
our own galaxy (see Figure~\ref{fig4}) and the error in the second ratio
does not allow us to break this degeneracy. So, although we obtain a
best-fit value that is consistent with our best extinction estimate, this
value is unconstrained and therefore not used.

The other methods used do not measure the extinction in the optical and
rely on correlations between various observed quantities and the optical
extinction to get an estimate. This necessary extrapolation renders these
techniques more uncertain. The correlations between hydrogen column density
and color excess (or extinction), and between the infrared 100 micron
emission and color excess are known to have large intrinsic scatters (e.g.,
\cite{boh78}; \cite{BP88}). The infrared emission has the additional
disadvantage that, near the galactic plane, it depends on the galactic
longitude, because the contribution of the stellar radiation field to the
dust emission declines with distance from the galactic center
(\cite{row91}).  GRB970228 is located near the galactic anticenter at
$l_{II}= 189.913\arcdeg$, $b_{II}=-17.941$, and it is conceivable that the
relation we have used to obtain the optical extinction underestimates
it. On the other hand, the value obtained by SDF98 is very similar and is
based on a different calibration that does not depend on this correlation,
so the effect of galactic longitude is unclear.  Another concern about
these methods is the spatial resolution. The resolution of the hydrogen
column density maps that we have used is of the order of a degree, while
the resolution of the IRAS 100 micron maps that we have used is
approximately 5 arcminutes. Figure~\ref{fig6} shows both maps. The two maps
correlate well on large angular scales. However, the resolution of the
hydrogen column density map is clearly insufficient for a good
determination of the extinction towards GRB970228 because there is a strong
gradient in this region. The IRAS 100 micron image, with its higher
resolution, demonstrates this point. Structures of a few arcminutes in size
that are visible in the 100 $\micron$ map are smeared out in the HI
image. It is important to note that the infrared emission is very knotty in
the GRB region that coincides with the outskirts of a SNR centered
$\sim10\arcdeg$ to the East (left) outside of our image in
Figure~\ref{fig6}. It is unclear, whether material from this SNR can
contribute to the extinction in the GRB field, although from the 100
$\micron$ image this seem unlikely. The IRAS 100 micron image does not
provide enough resolution for a definitive conclusion.

Table~\ref{tbl2} summarizes our measurements. We have computed the optical
extinction using three different methods. We have also used correlations
with neutral hydrogen and dust emission for additional estimates. The
measured X-ray spectrum of the GRB afterglow provides another measurement
of the extinction as well. Unfortunately, given the faintness of the GRB
X-ray afterglow when observed by BeppoSAX this constraint is very weak and
completely superseded by our measurements. We compute our best value for
the optical extinction with the weighted average of the values obtained by
the different methods. We apply two weights to average our
measurements. One is inversely proportional to the relative error and the
other is based on our subjective evaluation of the method reliability. We
arbitrarily assign a three times larger weight to methods based on optical
data than those based on other wavelengths. These weights are tabulated in
Table~\ref{tbl2} third and fourth column respectively.

\begin{figure}[t]
\columnwidth13cm
\plotone{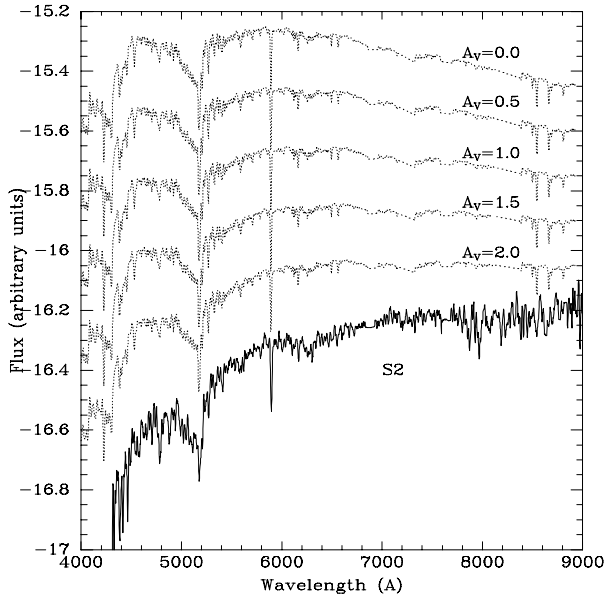}
\caption[]
{Spectrum of star S2 (solid spectrum) and a K4v star library spectrum
reddened by $A_V=0.0$, 0.5, 1.0, 1.5 and 2.0, respectively from top to
bottom (dotted spectra).}
\label{fig5b}
\end{figure}

It is somewhat puzzling that not all the methods used give estimates of the
extinction that are consistent with each other. The two methods based on
spectroscopic observations give higher values. This effect could be due to
a wrong spectrophotometric calibration. However, the agreement between
photometric and spectrophotometric colors (Table~\ref{tbl1}) indicates that
this is not the case. A spectral type misclassification could also yield to
an overestimation of the extinction using stellar spectra. We believe that
we have been very cautious assigning spectral types (see Section 3.2),
which results in the large errors quoted. For example, if the value of the
extinction is $A_V\sim0.9$, star S2 had to be an M star in order to be
compatible with the observed colors (or overall spectral slope) . The lack
of strong molecular features in the spectrum rules out this possibility. To
illustrate the scope of the disagreement we plot in Figure~\ref{fig5b} the
observed spectrum of star S2 along with a stellar library spectra of a K4v
star, that is reddened with an $R_V=3.1$ extinction law of $A_V=0.0$, 0.5,
1.0 1.5 and 2.0. A value of the extinction around $A_V\sim0.9$ would imply
a reddened K4v star that has a markedly different spectral slope than that
of star S2. If the extinction value measured using the stellar spectra is
at least $A_V\sim1.2$ (the best value obtained with the Balmer emission
line ratios agrees with such a figure), why do the other methods give a
significant lower value? We believe that a plausible explanation could be
variations in the extinction on angular scales $<1$ arcminute. The IRAS 100
$\micron$ map provides evidence of strong variations of the dust emission
on scales of a few arcminutes (Figure~\ref{fig6}). It is therefore likely
that the difference between extinction estimates derived from hydrogen and
infrared dust emissions and the extinction estimate derived from the
spectra of the nearby stars is due to variations on arcminute
scales. However, in order to explain the difference between the extinction
estimates obtained from galaxy number counts and the spectra of nearby
stars, those variations would have to be on scales $<1$ arcminutes. This
would have to be so because the WFPC2 images cover scales of the order of
one arcminutes, while stars S1 and S2, which have the most weight in the
estimation using stellar spectra, are only 2.9'' and 16.8'' arcseconds away
from GRB970228. Some support for the hypothesis of variations in the
extinction on small angular scales comes from the fact that the PC1
CCD has fewer objects than expected from extrapolation of the WF CCDs
counts; however, this deficit is not statistically significant.

\markcite{gon98}Gonz\'alez et al. (1998) also estimate the extinction in
this field. They compare cumulative galaxy number counts and galaxy colors
in the GRB970228 field to those in several other HST WFPC2 images to obtain
a value of the extinction. They obtain values consistent with the SFD98
estimate.

Summarizing, our best estimate for the optimal extinction is
$A_V=1.19^{+0.10}_{-0.17}$. Such a value considerably modifies the
broad-band spectral shape of the GRB970228 afterglow (\cite{rei98}).  For
example, an observed broad band $V-I_c$ color of 2.0 would deredden to an
unabsorbed $V-I_c=1.5$ and a $V$ magnitude of 22.0 would turn into
$V=20.8$. If we were to drop the determinations of the extinction from the
stellar spectra, we would obtain $A_V=0.86\pm0.11$, but we feel that there
is no a priori reason why these should be discarded.  We emphasize that the
stars lie close to and bracket the position of the GRB970228 afterglow.

\section{Conclusions}

We have measured the extinction towards the field of GRB970228. In making
this determination, we have used the relative number counts between the
WFPC2 images of the HDF and GRB970228 fields, the spectra of three nearby
stars and the Balmer line ratios of a nearby galaxy. The first method
produces the tightest constraint due to the relative good statistics. The
second method constitutes the best way of determining the extinction
because the widest spectral range is covered and the stars lie near the
direction of GRB970228. However, due to poor signal-to-noise the
constraints are poor. The third method gives an unconstrained value of the
extinction, although it is convenient to remark that its best value
coincides with our final result within the errors.  Variance among the
results of these methods may indicate that the gas and dust in this
direction are ``mottled'' or clumped on small angular scales.  We have used
other indirect methods based on the hydrogen column density and the dust
100$\micron$ emission. The infrared emission gives a lower estimate for the
extinction than the previous methods.  Combining the above techniques,
weighting the optical methods more than the other indirect ones, we obtain
a best value of $A_V=1.19^{+0.10}_{-0.17}$.

The measured $A_V$ implies that the GRB970228 afterglow is intrinsically
brighter and bluer than the observed magnitudes and color. For example, in
the $R_c$ filter the intrinsic magnitudes is 1.2 brighter and the $V-I_c$
color is 0.52 magnitudes bluer than observed.

\acknowledgments

It is a pleasure to acknowledge Carlo Graziani, Daniel Reichart and Jean
Quashnock, for their help, especially with regard to statistical
methodology. We have also benefited from useful discussions with Mark
Metzger, Cole Miller, Dave Cole and Andrew Fruchter. We thank Andrew
Fruchter in particular for pointing out to us the effect on galaxy number
counts of surface brightness dimming due to extinction. We thank John
Tonry, Esther Hu, Len Cowie and Richard McMahon for making publicly
available their spectra of the objects in the GRB97028 field. We have made
extensive use of the SkyView facility developed and maintained under NASA
grants at GSFC. Part of this work is based on NASA/ESA Hubble Space
Telescope archival data retrieved from the archive maintained at STSci. We
acknowledge support from NASA grants NAGW-4690, NAG 5-1454, and NAG 5-4406.


\clearpage

%
\begin{deluxetable}{ccccc}
\tablecaption{Star magnitudes.\label{tbl1}}
\tablewidth{12cm}
\tablehead{
\colhead{Star} & \colhead{$V_{606}$} & \colhead{$I_{814}$} & \colhead{$V_{606}-I_{814}$} & \colhead{$(V_{606}-I_{814})_{spec}$}
} 
\startdata
S1 & $22.63\pm0.02$ & $21.68\pm0.03$ & $0.95\pm0.04$ & $0.93$ \nl
S2 & $21.52\pm0.02$ & $20.63\pm0.03$ & $0.89\pm0.04$ & $0.93$ \nl
S3 & $24.14\pm0.06$ & $22.32\pm0.06$ & $1.82\pm0.09$ & $1.88$ \nl
\enddata
\end{deluxetable}

\begin{deluxetable}{lccc}
\tablecaption{Extinction values.\label{tbl2}}
\tablewidth{12cm}
\tablehead{
\colhead{Method} & \colhead{$A_V$} & \colhead{weight1} & \colhead{weight2}} 
\startdata
Number counts                 & $0.89\pm0.13$           & 0.357 & 0.375 \nl
Balmer series ratios          & $1.27$                  & 0.000 & 0.000 \nl
Stars                         & $1.73^{+0.20}_{-0.42}$  & 0.291 & 0.375 \nl
X-ray extinction              & $< 2.26$                & 0.000 & 0.000 \nl
$N(HI)$                       & $0.93\pm0.39$           & 0.124 & 0.125 \nl
$I_{100 \mu m}$ (correlation) & $0.79\pm0.21$           & 0.000 & 0.000 \nl
$I_{100 \mu m}$ (SFD98)       & $0.70\pm0.16$           & 0.228 & 0.125 \nl
\tableline
Combined (weight1)    & $1.10^{+0.10}_{-0.14}$ &  & \nl
Combined (weight1\&2) & $1.19^{+0.10}_{-0.17}$ &  & \nl
\enddata
\end{deluxetable}


%
%







\end{document}